\documentclass[twocolumn,showpacs,preprintnumbers,amsmath,amssymb,superscriptaddress]{revtex4}

\usepackage{graphicx}
\usepackage{dcolumn}
\usepackage{bm}

\begin{document}


\title{Direct observation of spin-polarized surface states in the parent compound of topological insulator Bi[1-x]Sb[x] using spin-resolved-ARPES in a 3D Mott-polarimetry spin mode}

\author{D. Hsieh}
\affiliation{Joseph Henry Laboratories of Physics, Princeton
University, Princeton, NJ 08544, USA}
\author{Y. Xia}
\affiliation{Joseph Henry Laboratories of Physics, Princeton
University, Princeton, NJ 08544, USA}
\author{L. Wray}
\affiliation{Joseph Henry Laboratories of Physics, Princeton
University, Princeton, NJ 08544, USA}
\author{D. Qian}
\affiliation{Joseph Henry Laboratories of Physics, Princeton
University, Princeton, NJ 08544, USA}
\author{J. H. Dil}
\affiliation{Physik-Institut, Universit\"{a}t Z\"{u}rich-Irchel,
8057 Z\"{u}rich, Switzerland} \affiliation{Swiss Light Source, Paul
Scherrer Institute, CH-5232, Villigen, Switzerland}
\author{F. Meier}
\affiliation{Physik-Institut, Universit\"{a}t Z\"{u}rich-Irchel,
8057 Z\"{u}rich, Switzerland} \affiliation{Swiss Light Source, Paul
Scherrer Institute, CH-5232, Villigen, Switzerland}
\author{L. Patthey}
\affiliation{Swiss Light Source, Paul Scherrer Institute, CH-5232,
Villigen, Switzerland}
\author{J. Osterwalder}
\affiliation{Physik-Institut, Universit\"{a}t Z\"{u}rich-Irchel,
8057 Z\"{u}rich, Switzerland}
\author{G. Bihlmayer}
\affiliation{Institut f\"{u}r Festk\"{o}rperforschung,
Forschungszentrum J\"{u}lich, , D-52425 J\"{u}lich, Germany}
\author{Y.S. Hor}
\affiliation{Department of Chemistry, Princeton University,
Princeton, NJ 08544, USA}
\author{R.J. Cava}
\affiliation{Department of Chemistry, Princeton University,
Princeton, NJ 08544, USA}
\author{M.Z. Hasan}
\affiliation{Joseph Henry Laboratories of Physics, Princeton
University, Princeton, NJ 08544, USA} \affiliation{Princeton Center
for Complex Materials, Princeton University, Princeton, NJ 08544,
USA}

\date{\today}

\begin{abstract}
We report high-resolution spin-resolved photoemission spectroscopy (Spin-ARPES) measurements on the parent compound Sb of the first discovered 3D topological insulator Bi$_{1-x}$Sb$_{x}$ [D. Hsieh \textit{et al.}, Nature \textbf{452}, 970 (2008) Submitted \textbf{2007}]. By modulating the incident photon energy, we are able to map both the bulk and (111) surface band structure, from which we directly demonstrate that the surface bands are spin polarized by the spin-orbit interaction and connect the bulk valence and conduction bands in a topologically non-trivial way. A unique asymmetric Dirac surface state gives rise to a $k$-splitting of its spin polarized electronic channels. These results complement our previously published works on this materials class and re-confirm our discovery of first bulk (3D) topological insulator - topological order in bulk solids. \end{abstract}

\maketitle

Topological insulators are a new phase of quantum matter that are theoretically distinguished from ordinary insulators by a $Z_2$
topological number that describes its bulk band structure
\cite{Kane, Moore, Fu_inversion}. They are characterized by a bulk
electronic excitation gap that is opened by spin-orbit coupling, and
unusual metallic states that are localized at the boundary of the
crystal. The quantum spin Hall insulator \cite{Bernevig, Konig}, is simply understood as two copies of the integer quantum Hall systems \cite{Haldane} where the spin-orbit coupling acts as a magnetic
field that points in a spin dependent direction, giving rise to
counter propagating spin polarized states on the 1D
crystal edge. Three-dimensional topological insulators on the other
hand have no quantum Hall analogue. Its surface states, which are
necessarily spin polarized, realize a novel 2DEG metal that remains
delocalized even in the presence of disorder due to its spin-momentum locking \cite{Fu_3D,Fu_inversion, Moore, Murakami}. For these reasons, they have also been proposed as a route to dissipationless spin currents
which, unlike current semiconductor heterostructure based spintronics devices, do not require an externally applied electric field.

\begin{figure}[h]
\includegraphics[scale=0.4,clip=true, viewport=0.0in 0in 8.8in 10.5in]{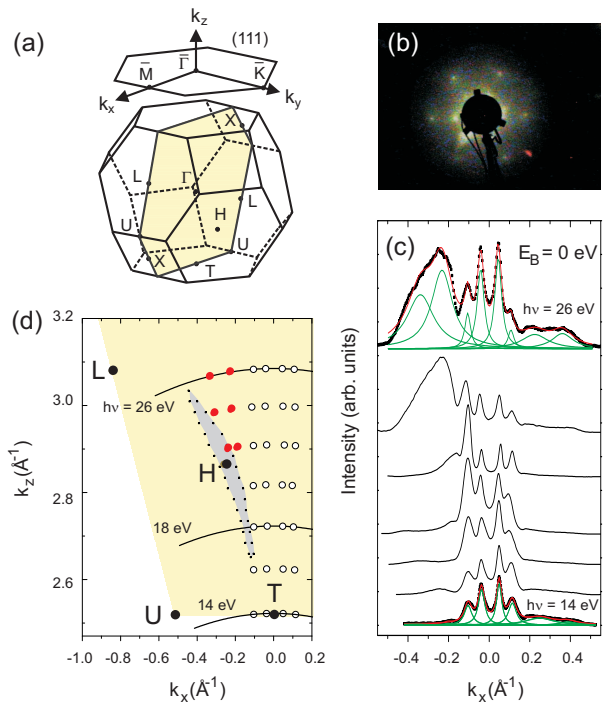}
\caption{\label{fig:Sb_Fig1} Experimental separation of bulk from
surface electron states in Sb using ARPES. (a) Schematic of the bulk
BZ of Sb and its (111) surface BZ. The shaded region denotes the
momentum plane in which the following ARPES spectra were measured.
(b) LEED image of the \textit{in situ} cleaved (111) surface
exhibiting a clear hexagonal symmetry. (c) Select MDCs at the Fermi
level taken with photon energies from 14 eV to 26 eV in steps of 2
eV, taken in the $TXLU$ momentum plane. Peak positions in the MDCs
were determined by fitting to Lorentzians (red curves). (d)
Experimental 3D bulk Fermi surface near H (red circles) and 2D
surface Fermi surface near $\bar{\Gamma}$ (open circles) determined
by matching the fitted peak positions from (c) to calculated
constant $h\nu$ contours (black curves). Theoretical hole Fermi
surface calculated in \cite{Falicov}.}
\end{figure}

Recent photoemission \cite{Hsieh, HsiehSci} suggest that single crystals of insulating Bi$_{1-x}$Sb$_{x}$ ($0.07\leq x \leq0.22$) alloys realize
a 3D topological insulator. The non-trivial $Z_2$ invariant that
characterizes Bi$_{1-x}$Sb$_{x}$ is inherited from the bulk band
structure of pure Sb \cite{Fu_inversion, Murakami}, therefore,
although Sb is a bulk semimetal, its non-trivial bulk band topology
should be manifest in its surface state spectrum. Such a study
requires a separation of the Fermi surface of the surface states of
Sb from that of its bulk states over the entire surface Brillouin
zone (BZ), as well as a direct measurement of the spin degeneracy of
the surface states. To date, angle-resolved photoemission
spectroscopy (ARPES) experiments on low lying states have only been
performed on single crystal Sb with fixed He I$\alpha$ radiation,
which does not allow for separation of bulk and surface states
\cite{Sugawara}. Moreover the aforementioned study, as well as ARPES
experiments on Sb thin films \cite{Hochst_Sbfilm}, only map the band
dispersion near $\bar{\Gamma}$, missing the band structure near
\={M} that is critical to determining the $Z_2$ invariant
\cite{Hsieh}. In this Letter, we have performed spin- and
angle-resolved photoemission experiments on single crystal Sb(111).
Using variable photon energies, we successfully isolate the surface
from bulk electronic bands over the entire BZ and map them with spin
sensitivity. We show directly that the surface states are gapless
and spin split, and that they connect the bulk valence and
conduction bands in a topologically non-trivial way.

Spin-integrated ARPES measurements were performed with 14 to 30 eV
photons on beam line 5-4 at the SSRL. Spin resolved ARPES
measurements were performed at the SIS beam line at the SLS using
the COPHEE spectrometer \cite{Hoesch2} with a single 40 kV classical
Mott detector and a photon energy of 20 eV. The typical energy and
momentum resolution was 15 meV and 1\% of the surface BZ
respectively at beam line 5-4, and 80 meV and 3\% of the surface BZ
respectively at SIS using a pass energy of 3 eV. High quality single
crystals of Sb and Sb$_{0.9}$Bi$_{0.1}$ were grown by methods
detailed in \cite{Hsieh, HsiehSci}. Cleaving these samples \textit{in situ} between 10 K and 55 K at chamber pressures less than 5
$\times10^{-11}$ torr resulted in shiny flat surfaces, characterized
by low energy electron diffraction to be clean and well ordered with
the same symmetry as the bulk [Fig. ~\ref{fig:Sb_Fig1}(a) \& (b)].
This is consistent with photoelectron diffraction measurements that
show no substantial structural relaxation of the Sb(111) surface
\cite{Bengio}. Band calculation was performed using the full
potential linearized augmented plane wave method in film geometry as
implemented in the FLEUR program and local density approximation for
description of the exchange correlation potential \cite{Koroteev}.

Figure ~\ref{fig:Sb_Fig1}(c) shows momentum distributions curves
(MDCs) of electrons emitted at $E_F$ as a function of $k_x$
($\parallel$ $\bar{\Gamma}$-\={M}) for Sb(111). The out-of-plane
component of the momentum $k_z$ was calculated for different
incident photon energies ($h\nu$) using the free electron final
state approximation with an experimentally determined inner
potential of 14.5 eV \cite{Hochst_Sbfilm}. There are four peaks in
the MDCs centered about $\bar{\Gamma}$ that show no dispersion along
$k_z$ and have narrow widths of $\Delta k_x \approx$ 0.03
\AA$^{-1}$. These are attributed to surface states and are similar
to those that appear in Sb(111) thin films \cite{Hochst_Sbfilm}. As
$h\nu$ is increased beyond 20 eV, a broad peak appears at $k_x
\approx$ -0.2 \AA$^{-1}$, outside the $k$ range of the surface
states near $\bar{\Gamma}$, and eventually splits into two peaks.
Such a strong $k_z$ dispersion, together with a broadened linewidth
($\Delta k_x \approx$ 0.12 \AA$^{-1}$), is indicative of bulk band
behavior, and indeed these MDC peaks trace out a Fermi surface [Fig.
~\ref{fig:Sb_Fig1}(d)] that is similar in shape to the hole pocket
calculated for bulk Sb near H \cite{Falicov}. Therefore by choosing
an appropriate photon energy (e.g. $\leq$ 20 eV), the ARPES spectrum
along $\bar{\Gamma}$-\={M} will have contributions from only the
surface states. The small bulk electron pocket centered at L is not
accessed using the photon energy range we employed [Fig.
~\ref{fig:Sb_Fig1}(d)].

\begin{figure}
\includegraphics[scale=0.31,clip=true, viewport=0.0in 0in 11.5in 8.0in]{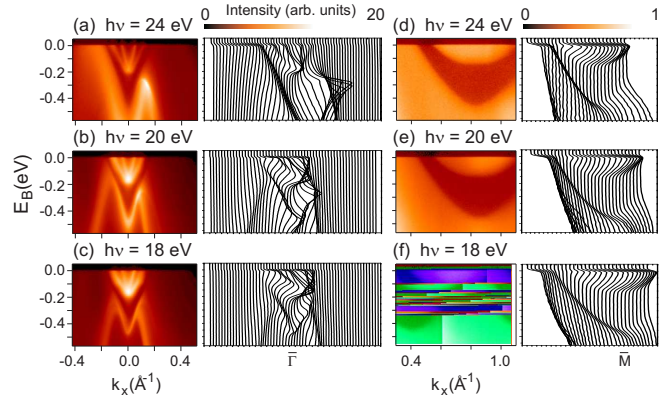}
\caption{\label{fig:Sb_Fig2} Surface and bulk band dispersion. ARPES
intensity maps as a function of $k_x$ near $\bar{\Gamma}$ (a)-(c)
and \={M} (d)-(f) and their corresponding EDCs, taken using $h\nu$ =
24 eV, 20 eV and 18 eV photons. The intensity scale of (d)-(f) is a
factor of about twenty smaller than that of (a)-(c) due to the
intrinsic weakness of the ARPES signal near \={M}.}
\end{figure}

\begin{figure}
\includegraphics[scale=0.41,clip=true, viewport=0.0in 0in 11.0in 9.8in]{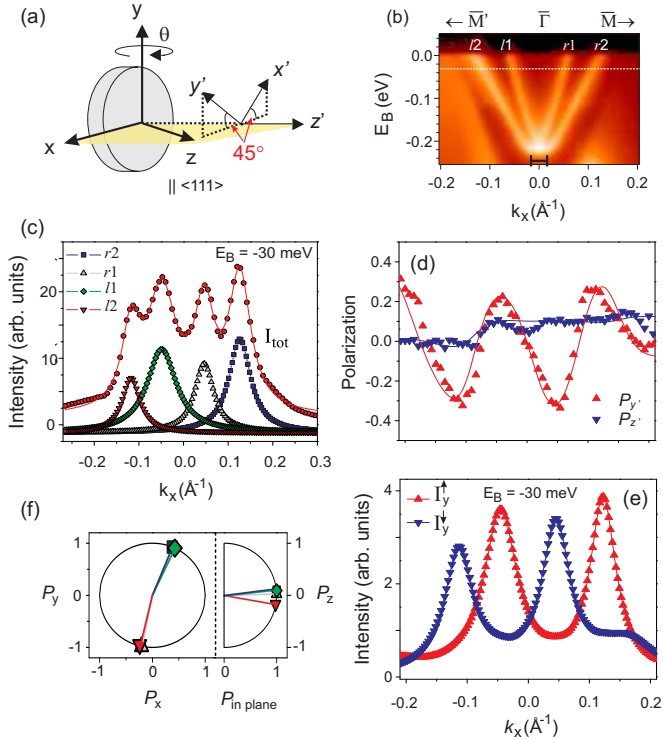}
\caption{\label{fig:Sb_Fig3} Large spin splitting of surface states
on Sb(111). (a) Experimental geometry of the spin resolved ARPES
study. At normal emission ($\theta$=0$^{\circ}$), the sensitive
$y'$-axis of the Mott detector is rotated by 45$^{\circ}$ from the
sample $\bar{\Gamma}$-\={M} ($\parallel$ x) direction, and the
sensitive $z'$-axis of the Mott detector is parallel to the sample
normal. Spin up and down are measured with respect to these two
quantization axes. (b) Spin integrated ARPES spectra along the
\={M'}-$\bar{\Gamma}$-\={M} direction taken using a photon energy
$h\nu$ = 22 eV. The momentum splitting between the band minima is
indicated by the black bar and is approximately 0.03\AA$^{-1}$. (c)
Momentum distribution curve of the spin integrated spectra at $E_B$
= -30 meV (shown in (b) by white line) using a photon energy $h\nu$
= 20 eV, together with the Lorentzian peaks of the fit. (d) Measured
spin polarization curves (symbols) for the $y'$ and $z'$ components
together with the fitted lines using the two-step fitting routine.
Even though the measured polarization only reaches a magnitude of
around $\pm$0.4, similar to what is observed in thin film Bi(111)
\cite{Hirahara}, this is due to a non-polarized background and
overlap of adjacent peaks with different spin polarization. The
fitted parameters are in fact with consistent with 100\% polarized
spins. (e) Spin resolved spectra for the y component based on the
fitted spin polarization curves shown in (d). (f) The in-plane and
out-of-plane spin polarization components in the sample coordinate
frame obtained from the spin polarization fit. The symbols refer to
those in (c).}
\end{figure}

\begin{figure}
\includegraphics[scale=0.32,clip=true, viewport=0.0in 0in 11.0in 6.9in]{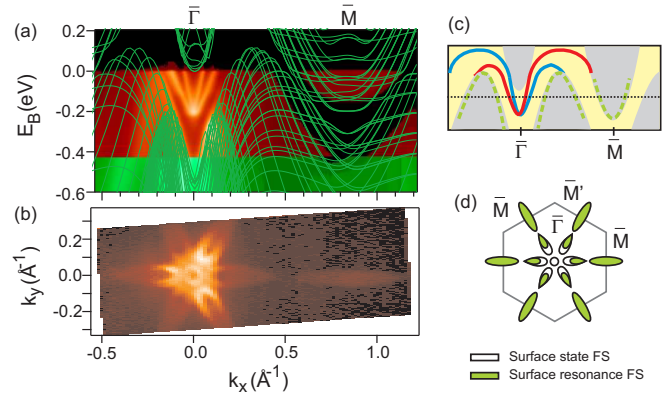}
\caption{\label{fig:Sb_Fig4} Topologically non-trivial surface
states of Sb(111). (a) Calculated surface state band structure for
freestanding 20 bilayer Sb(111) slabs together with an ARPES
intensity map of Sb(111) along the $\bar{\Gamma}$-\={M} direction
taken with $h\nu$ = 22 eV photons. Green curves show the calculated
bulk bands along the $k_x$ direction projected onto the (111) plane.
(b) ARPES intensity map at $E_F$ in the $k_x$-$k_y$ plane taken with
$h\nu$ = 20 eV photons. (c) Schematic picture showing that the
gapless spin polarized surface bands (red and blue lines) connect
the projected bulk valence and conduction bands (shaded regions) and
are thus topologically non-trivial. The surface resonances (dashed
green lines) do not connect the bulk valence and conduction bands
and are thus topologically trivial. (d) Schematic of the surface
Fermi surface topology of Sb(111) showing the pockets formed by the
pure surface states (unfilled) and the surface resonances (filled
green). The purely surface state Fermi contours enclose only the one
surface TRIM located at $\bar{\Gamma}$.}
\end{figure}

ARPES spectra along $\bar{\Gamma}$-\={M} taken at three different
photon energies are shown in Fig. ~\ref{fig:Sb_Fig2}. Near
$\bar{\Gamma}$ there are two rather linearly dispersive electron
like bands that meet exactly at $\bar{\Gamma}$ at a binding energy
$E_B \sim$ -0.2 eV. This behavior is consistent with previous ARPES
measurements along the $\bar{\Gamma}$-\={K} direction
\cite{Sugawara} and is thought to come from a pair of spin-split
surface bands that become degenerate at the time reversal invariant
momentum (TRIM) $\bar{\Gamma}$ due to Kramers degeneracy. The Fermi
velocities of the inner and outer V-shaped bands are $4.4 \pm 0.1$
eV$\cdot$\AA and $2.2 \pm 0.1$ eV$\cdot$\AA respectively as found by
fitting straight lines to their MDC peak positions. The surface
origin of this pair of bands is established by their lack of
dependence on $h\nu$ [Fig. ~\ref{fig:Sb_Fig2}(a)-(c)]. A strongly
photon energy dispersive hole like band is clearly seen on the
negative $k_x$ side of the surface Kramers pair, which crosses $E_F$
for $h\nu=$ 24 eV and gives rise to the bulk hole Fermi surface near
H [Fig. ~\ref{fig:Sb_Fig1}(d)]. For $h\nu\leq$ 20 eV, this band
shows clear back folding near $E_B \approx$ -0.2 eV indicating that
it has completely sunk below $E_F$. Further evidence for its bulk
origin comes from its close match to band calculations [Fig.
~\ref{fig:Sb_Fig2}(a)]. Interestingly, at photon energies such as 18
eV where the bulk bands are far below $E_F$, there remains a uniform
envelope of weak spectral intensity near the Fermi level in the
shape of the bulk hole pocket seen with $h\nu$ = 24 eV photons,
which is symmetric about $\bar{\Gamma}$. This envelope does not
change shape with $h\nu$ suggesting that it is of surface origin.
Due to its weak intensity relative to states at higher binding
energy, these features cannot be easily seen in the energy
distribution curves (EDCs) in Fig. ~\ref{fig:Sb_Fig2}(a)-(c), but
can be clearly observed in the MDCs shown in Fig.
~\ref{fig:Sb_Fig1}(c) especially on the positive $k_x$ side.
Centered about the \={M} point, we also observe a crescent shaped
envelope of weak intensity that does not disperse with $k_z$ [Fig.
~\ref{fig:Sb_Fig2}(d)-(f)], pointing to its surface origin. Unlike
the sharp surface states near $\bar{\Gamma}$, the peaks in the EDCs
of the feature near \={M} are much broader ($\Delta E \sim$80 meV)
than the spectrometer resolution (15 meV). The origin of this
diffuse ARPES signal is not due to surface structural disorder
because if that were the case, electrons at $\bar{\Gamma}$ should be
even more severely scattered from defects than those at \={M}. In
fact, the occurrence of both sharp and diffuse surface states
originates from a $k$ dependent coupling to the bulk as discussed
later.

To extract the spin polarization vector of each of the surface bands
near $\bar{\Gamma}$, we performed spin resolved MDC measurements
along the \={M}'-$\bar{\Gamma}$-\={M} cut at $E_B$ = -30 meV for
maximal intensity, and used the two-step fitting routine developed
in \cite{Meier}. The Mott detector in the COPHEE instrument is
mounted so that at normal emission it is sensitive to a purely
out-of-plane spin component ($z'$) and a purely in-plane ($y'$) spin
component that is rotated by 45$^{\circ}$ from the sample
$\bar{\Gamma}$-\={M} direction [Fig. ~\ref{fig:Sb_Fig3}(a)]. Each of
these two directions represents a normal to a scattering plane,
defined by the electron incidence direction on a gold foil and two
detectors mounted on either side that measure the left-right
asymmetry
$A_{y',z'}=[(I^{y',z'}_{L}-I^{y',z'}_{R})/(I^{y',z'}_{L}+I^{y',z'}_{R})]$
of electrons backscattered off the gold foil \cite{Hoesch2}. Figure
~\ref{fig:Sb_Fig3}(d) shows the spin polarization for both
components given by $P=(1/S_{eff})\times A^{y',z'}$, where
$S_{eff}=0.085$ is the Sherman function. Following the procedure
described in \cite{Meier}, we take the spins to be fully polarized,
assign a spin resolved spectra for each of the fitted peaks $I^{i}$
shown in Fig. ~\ref{fig:Sb_Fig3}(c), and fit the calculated
polarization spectrum to measurement. The spin resolved spectra for
the y component derived from the polarization fit is shown in Fig.
~\ref{fig:Sb_Fig3}(e), given by
$I_{y}^{\uparrow,\downarrow}=\sum_{i=1}^{4}I^{i}(1\pm
P_{y}^{i})/6+B/6$, where $B$ is a background and $P_{y}^{i}$ is the
fitted y component of polarization. There is a clear difference in
$I_{y}^{\uparrow}$ and $I_{y}^{\downarrow}$ at each of the four MDC
peaks indicating that the surface state bands are spin polarized.
Each of the pairs $l2/l1$ and $r1/r2$ have opposite spin, consistent
with the behavior of a spin split Kramers pair, and the spin
polarization of these bands are reversed on either side of
$\bar{\Gamma}$ in accordance with time reversal symmetry [Fig.
~\ref{fig:Sb_Fig3}(f)]. Similar to Au(111) \cite{Hoesch} and
W(110)-(1$\times$1)H \cite{Hochstrasser}, the spin polarization of
each band is largely in-plane consistent with a predominantly
out-of-plane electric field at the surface. However unlike the case
in Au(111), where the surface band dispersion is free electron like
and the magnitude of the Rashba coupling can be quantified by the
momentum displacement between the spin up and spin down band minima
\cite{Hoesch}, the surface band dispersion of Sb(111) is highly
non-parabolic. A comparison of the $k$-separation between spin split
band minima near $\bar{\Gamma}$ of Sb(111) [Fig.
~\ref{fig:Sb_Fig3}(b)] with those of Bi(111) \cite{Koroteev}, which
are 0.03 \AA$^{-1}$ and 0.08 \AA$^{-1}$ respectively, nevertheless
are consistent with the difference in their atomic $p$ level
splitting of Sb(0.6 eV) and Bi(1.5 eV) \cite{Liu}.

Figure ~\ref{fig:Sb_Fig4}(a) shows the full ARPES intensity map from
$\bar{\Gamma}$ to \={M} together with the calculated bulk bands of
Sb projected onto the (111) surface. Although the six-fold
rotational symmetry of the surface band dispersion is not known $a$
$priori$ due to the three-fold symmetry of the bulk, we measured an
identical surface band dispersion along $\bar{\Gamma}$-\={M'}. The
spin-split Kramers pair near $\bar{\Gamma}$ lie completely within
the gap of the projected bulk bands near $E_F$ attesting to their
purely surface character. In contrast, the weak diffuse hole like
band centered near $k_x$ = 0.3 \AA$^{-1}$ and electron like band
centered near $k_x$ = 0.8 \AA$^{-1}$ lie completely within the
projected bulk valence and conduction bands respectively. Thus
their ARPES spectra exhibit the expected lifetime broadening due to
hybridization with the underlying bulk continuum \cite{Kneedler}, a characteristic of surface resonance states. Figure
~\ref{fig:Sb_Fig4}(b) shows the ARPES intensity plot at $E_F$ of
Sb(111) taken at a photon energy of 20 eV, where the bulk band near
H is completely below $E_F$ [Fig. ~\ref{fig:Sb_Fig2}(b)]. Therefore
this intensity map depicts the topology of the Fermi surface due
solely to the surface states. By comparing Figs
~\ref{fig:Sb_Fig4}(a) and (b), we see that the inner most spin
polarized V-shaped band produces the circular electron Fermi surface
enclosing $\bar{\Gamma}$ while the outer spin polarized V-shaped
band produces the inner segment (0.1 \AA$^{-1}\leq$ $k_x\leq$ 0.15
\AA$^{-1}$) of the six hole Fermi surfaces away from $\bar{\Gamma}$.
Previous ARPES experiments along the $\bar{\Gamma}$-\={K} direction
\cite{Sugawara} show that this outer V-shaped band merges with the
bulk valence band, however the exact value of $k_x$ where this
occurs along the $\bar{\Gamma}$-\={M} direction is unclear since
only occupied states are imaged by ARPES. The outer segment of the
six hole pockets is formed by the hole like surface resonance state
for 0.15 \AA$^{-1}\leq$ $k_x\leq$ 0.4 \AA$^{-1}$. In addition, there
are electron Fermi surfaces enclosing \={M} and \={M'} produced by
surface resonance states at the BZ boundaries. Altogether, these
results show that in a single surface BZ, the bulk valence and
conduction bands are connected by a lone Kramers pair of surface
states [Fig.~\ref{fig:Sb_Fig4}(c)].

\begin{figure}
\includegraphics[scale=0.32,clip=true, viewport=0.0in 0in 11.0in 5.0in]{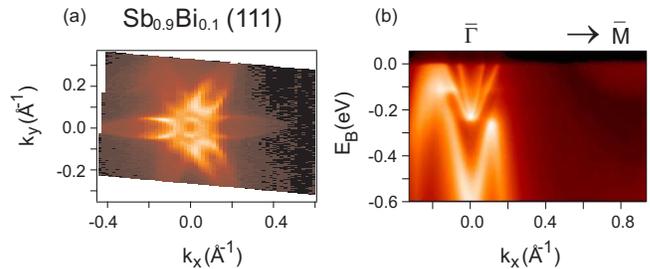}
\caption{\label{fig:Sb_Fig5} Spin split surface states survive
alloying disorder in Sb$_{0.9}$Bi$_{0.1}$. (a) ARPES intensity map
at $E_F$ of single crystal Sb$_{0.9}$Bi$_{0.1}$(111) in the
$k_x$-$k_y$ plane taken using 20 eV photons. (b) ARPES intensity map
of Sb$_{0.9}$Bi$_{0.1}$(111) along the $\bar{\Gamma}$-\={M}
direction taken with $h\nu$ = 22 eV photons.}
\end{figure}

In general, the spin degeneracy of surface bands on spin-orbit
coupled insulators can be lifted due to the breaking of space
inversion symmetry. However Kramers theorem requires that they
remain degenerate at four special time reversal invariant momenta
(TRIM) on the 2D surface BZ, which for Sb(111) are located at
$\bar{\Gamma}$ and three \={M} points rotated by 60$^{\circ}$ from
one another. According to recent theory, there are a total of four
$Z_2$ topological numbers $\nu_0$;($\nu_{1}\nu_{2}\nu_{3}$) that
characterize a 3D spin-orbit coupled insulator's bulk band structure
\cite{Fu_3D, Moore}. One in particular ($\nu_0$) determines
whether the spin polarized surface bands cross $E_F$ an even or odd
number of times between any pair of surface TRIM, and consequently
whether the insulator is trivial ($\nu_0$=0) or topological
($\nu_0$=1). An experimental signature of topologically non-trivial
surface states in insulating Bi$_{1-x}$Sb$_x$ is that the spin
polarized surface bands traverse $E_F$ an odd number of times
between $\bar{\Gamma}$ and \={M} \cite{Hsieh, Teo, Fu_inversion}.
Although this method of counting cannot be applied to Sb because it
is a semimetal, since there is a direct gap at every bulk $k$-point,
it is meaningful to assume some perturbation, such as alloying with
Bi \cite{Lenoir} that does not significantly alter the spin
splitting [Fig.~\ref{fig:Sb_Fig5}], that pushes the bulk valence H
and conduction L bands completely below and above $E_F$ respectively
without changing its $Z_2$ class. Under such an operation, it is
clear that the spin polarized surface bands must traverse $E_F$ an
odd number of times between $\bar{\Gamma}$ and \={M}, consistent
with the 1;(111) topological classification of Sb. This conclusion
can also be reached by noticing that the spin-split pair of surface
bands that emerge from $\bar{\Gamma}$ do not recombine at \={M},
indicative of a ``partner switching" \cite{Fu_3D} characteristic of
topological insulators.

In conclusion, we have mapped the spin structure of the surface
bands of Sb(111) and shown that the purely surface bands located in
the projected bulk gap are spin split by a combination of spin-orbit
coupling and a loss of inversion symmetry at the crystal surface.
The spin polarized surface states have an asymmetric Dirac like
dispersion that gives rise to its $k$-splitting between spin
up and spin down bands at $E_F$. Despite a modest atomic spin-orbit
coupling strength. This property of Sb in combination with its small
density of spin degenerate bulk states at the Fermi level due to its
semimetallic nature make it a promising candidate for high
temperature spin current sources. Moreover its topologically
non-trivial surface band structure makes Sb(111) an especially
appealing candidate for an unusual 2D Dirac protected free fermion system that exhibits antilocalization \cite{Fu_3D}.

\vspace{0.5cm}




\end{document}